\documentclass[twocolumn,aip]{revtex4} 

\usepackage{dsfont,amsmath,ulem,epsfig,stmaryrd,color}

\renewcommand{\Im}{{\rm Im}}
\newcommand{\ri}{{\rm i}}
\newcommand{\rd}{{\rm d}}
\newcommand{\rs}{{\rm s}}
\newcommand{\rp}{{\rm p}}
\newcommand{\re}{{\rm e}}

\newcommand{\Tr}{{\rm Tr}}

\begin{document}

\title{Local density of states above a disk --- geometrical vs. thermal boundary conditions}

\author{Svend-Age Biehs}
\email{ s.age.biehs@uni-oldenburg.de} 
\affiliation{Institut f{\"u}r Physik, Carl von Ossietzky Universit{\"a}t, D-26111 Oldenburg, Germany}
\affiliation{Center for Nanoscale Dynamics  (CeNaD), Carl von Ossietzky Universit\"{a}t, D-26129 Oldenburg, Germany}
\affiliation{Laboratoire Charles Coulomb (L2C), UMR 5221 CNRS-Université de Montpellier, F-34095 Montpellier, France}

\author{Achim Kittel}
\affiliation{Institut f{\"u}r Physik, Carl von Ossietzky Universit{\"a}t, D-26111 Oldenburg, Germany}
\affiliation{Center for Nanoscale Dynamics  (CeNaD), Carl von Ossietzky Universit\"{a}t, D-26129 Oldenburg, Germany}

\author{ Zhenghua An}
\affiliation{State Key Laboratory of Surface Physics and Institute for Nanoelectronic Devices and Quantum Computing, Department of Physics, Fudan University, Shanghai 200433, PR China}

\begin{abstract}
We analytically calculate the contribution to the local density of states due to thermal sources in a disk-like patch within the framework of fluctuational electrodynamics. We further introduce a wavevector cutoff method to approximate this contribution. We compare the results obtained with the source and cutoff method with the numerical exact LDOS above a metal disk attained by SCUFF-EM calculations. By this comparison we highlight the difference and resemblance of thermal and geometrical boundary conditions which are both relevant for near-field scanning microscope measurements. Finally, we give an outlook to general lateral temperature profiles and compare it with surface profiles.  
\end{abstract}

\maketitle

\section{Introduction}

In the last two decades, different kinds of scanning thermal microscopes have been developed which enable us to image the thermal near-field of solid interfaces in the infrared-region. A first near-field scanning thermal microscope of such kind has been set up in the research group of Yannik De Wilde~\cite{DeWilde2006,Babuty2013,Peragut2018}. This so-called Thermal Radiation Scanning Tunneling Microscope (TRSTM) is in principle an s-SNOM, which works without any external illumination. Therefore, it scatters the thermal near-field of a heated sample at the apex of the sharp TRSTM tip into the far field. The far-field signal can be decomposed into its different frequency parts, so that the TRSTM makes it possible to measure spectra of the thermal near-field in the vicinity of a sample. In order to obtain signals which are large enough to be measurable it is typically necessary to heat the samples by several hundreds of Kelvins. A similar AFM based near-field scanning thermal microscope, the so-called Thermal Infrared Near-field Spectroscope (TINS) has been set up in the group of Markus Raschke~\cite{Jones2012,Callahan2016}. A difference between the TINS and the TRSTM relies in the fact, that for TINS also the tip of the microscope can be heated. The far-field signal emitted and scattered by the heated tip can again be decomposed in its frequency components using FTIR. Thus, TINS and TRSTM allow for measuring the spectral information of a given sample. 

A third near-field scanning thermal microscope, the so-called Scanning Noise Microscope (SNoiM) has been developed by Susumu Komiyama and has been advanced by the group of An Zhenghua~\cite{Kajihara2010,Weng2018,Komiyama2019}. As for the TRSTM the SNoiM is in principle an s-SNOM without external illlumination.The important difference between the SNoiM and the TRSTM relies in the fact that the SNoiM has a ultra-sensitive single photon detector~\cite{An1, An2, Komiyama2011} which is working at cryogenic temperatures of 4.2K. Due to this specific detector, even very weak far-field signals can be measured such that for the SNoiM measurement it is not necessary to heat neither the microscope tip nor the sample as for the TRSTM or TINS. Consequently, SNoiM is used to measure signals of the microscope and the sample at room temperature. However, in contrast to the TRSTM or TINS the actual SNoiM setup can only measure signals at a single wavelength which is typically about 14.5~$\mu$m. 

Yet another near-field scanning thermal microscope is the Near-field Scanning Thermal Microscope (NSthM) designed by Achim Kittel~\cite{Kittel2005,Kittel2008,Worbes2013,Kloppstech2017}. It is in principle an STM which has been augmented by a thermocouple at the foremost part of the probe. In contrast to the TRSTM, TINS and SNoiM the NSThM measures the radiative heat flux between the probe and a cooled sample. The STM ability of the NSThM allows for precisely controlling the distances between the tip and the sample so that heat fluxes down to 0.3~nm distance can be measured. Like for the TRSTM, TINS and SNoiM the tip of the NSThM is very sharp having a tip radius of about 20~nm. Consequently, it is possible to acquire highly resolved images of the near-field thermal radiation. With the NSThM only spectrally integrated near-field heat fluxes can be measured so that the NSThM has no access to the spectrum of the heat flux. However, the NSThM has the advantage over the other microscopes that with its STM ability also topographical information of the sample can be obtained with a high lateral and vertical resolution.

The first theoretical models for the TRSTM, TINS, SNoiM and NSThM were based on the assumption that the foremost part of the probes of the microscopes can be regarded as small spheres which can be described as dipoles in the long-wavelength regime~\cite{Dorofeyev1998,Joulain2003,Dorofeyev2008,Biehs2008,Joulain2014,Jarzembski2017,Herz2018, Herz2021}. Therefore, the signals should in lowest order be proportional to the photonic local density of states (LDOS) at the position of the microscope tip and consequently the measured signals have been compared with the LDOS~\cite{Kittel2008,Babuty2013,Jones2012,Callahan2016,Weng2018}. Improved theoretical descriptions have been brought forward recently by a discrete-dipole model of the tip and an exact boundary-element method~\cite{Edalatpour2016,Kloppstech2017,Edalatpour2019,Herz2022}.

\begin{figure}
  \includegraphics[width=0.45\textwidth]{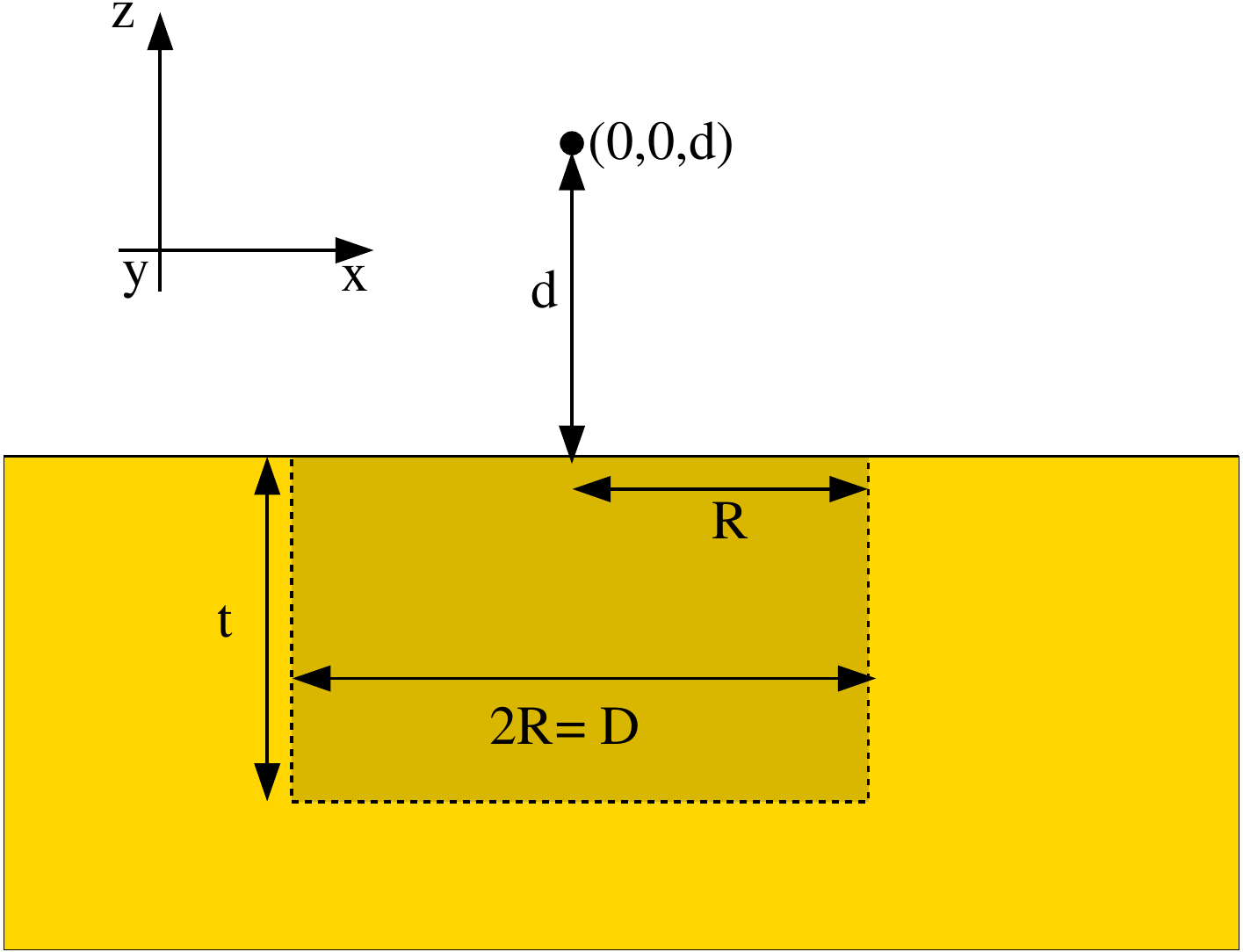}
  \caption{\label{Fig:sketchvolume} Sketch of the considered disk-like volume in a bulk with depth $t$ and diameter $D = 2 R$ (shaded part) in which the source currents contributing to the LDOS at the position $\mathbf{r} = (0,0,d)^t$ are considered.}
\end{figure}

\begin{figure}
  \includegraphics[width=0.45\textwidth]{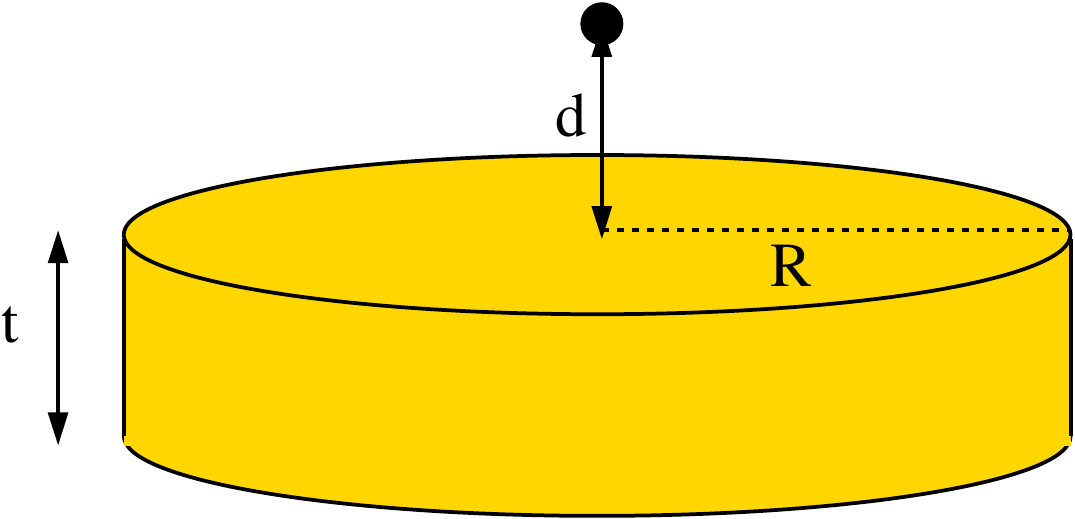}
  \caption{\label{Fig:sketch} Sketch of a gold nanodisk of thickness $t$ and with diameter $D = 2R$. The SNoiM has measured the LDOS at the distance $z$ above the center of the disk for disks with different diameters.}
\end{figure}

The aim of our work is to shed some light on the impact of geometrical and thermal boundary conditions on the LDOS which are both highly relevant for the above mentioned near-field thermal imaging methods. To this end, we derive an analytical expression for the contribution to the LDOS above a semi-infinite medium stemming from thermal sources in a disk-shaped patch as sketched in Fig.~\ref{Fig:sketchvolume} within the framework of fluctuational electrodynamics as done in Ref.~\cite{Zheng2014} for the van der Waals forces and the near-field radiative heat transfer. This LDOS would correspond to the signal of a SNoiM, for instance, for a planar sample which is heated in a disk-shaped area. We further numerically calculate the LDOS above a nanodisk as sketched in Fig.~\ref{Fig:sketch} by using SCUFF-EM~\cite{SCUFF1,SCUFF2}. This LDOS would correspond to a measurement of the LDOS above an free-standing nanodisk as conducted recently with a SNoiM~\cite{SNOIMPatch}. We compare the LDOS obtained with the two methods and discuss the similarities and differences of the impact of the thermal and geometrical boundary conditions. Furthermore, we introduce a simple wave-vector cutoff approximation as used in Ref.~\cite{SNOIMPatch,KomiyamaCutoff} and discuss its ability to mimic the LDOS calculated with the source method. Finally, we debate the relation between general lateral temperature and surface profiles.  

\section{Definition of the LDOS}

The electric and magnetic part $D_e$ and $D_m$ of the LDOS at a position $\mathbf{r}$ are defined via the electric and magnetic energy density of the electromagnetic field generated by the fluctuational sources inside the semi-infinite body via the relations~\cite{Agarwal,Eckhardt,Dorofeyev}  
\begin{align}
	u_{e} &= \frac{\epsilon_0}{2} \langle E_{\alpha} (\mathbf{r}) E_{\alpha} (\mathbf{r}) \rangle = \int_0^\infty \!\!\! \rd \omega \, \Theta(T) D_{e}(\omega,d), \label{Eq:Ue} \\
  u_{m} &= \frac{\mu_0}{2} \langle H_{\alpha} (\mathbf{r}) H_{\alpha} (\mathbf{r}) \rangle = \int_0^\infty \!\!\! \rd \omega \, \Theta(T) D_{m}(\omega,d) 
\end{align}
with 
\begin{equation}
  \Theta(T) = \frac{\hbar \omega}{\exp(\hbar \omega / k_{\rm B} T) - 1}
\end{equation} 
assuming that the thermal sources in the sample are in local equilibrium at a temperature $T$. $k_{\rm B}$ and $\hbar$ are the Boltzmann and the reduced Planck constant and $\epsilon_0/\mu_0$ are the permittivity and permeability of vacuum. From Rytov's fluctuational electrodynamics it follows for a homogeneous and isotropic material with permittivity $\epsilon_1$ that (Einstein convention)
\begin{align}
  D_e (\omega,d) &= \frac{k_0^3\epsilon_1''}{\pi c}  \int_V \!\!\! \rd^3 \mathbf{r}' \,\Tr\bigl[ \mathds{G}^{\rm EE} (\mathbf{r},\mathbf{r}') {\mathds{G}^{\rm EE}}^\dagger (\mathbf{r},\mathbf{r}') \bigr], \\
  D_m (\omega,d) &= \frac{k_0^3\epsilon_1''}{\pi c} \frac{\mu_0}{\epsilon_0} \int_V \!\!\! \rd^3 \mathbf{r}' \,  \Tr\bigl[ \mathds{G}^{\rm HE} (\mathbf{r},\mathbf{r}') {\mathds{G}^{\rm HE}}^\dagger (\mathbf{r},\mathbf{r}') \bigr]
\end{align}
where $ \mathds{G}^{\rm EE}, \mathds{G}^{\rm HE}$ are the dyadic Green's function of the electric and magnetic field, resp., $\epsilon_1''$ is the imaginary part of the permittivity and $k_0 = \omega/c$ is the wavenumber in vacuum and $c$ is the vacuum light velocity. The volume integral has to be taken over the volume $V$ which contains the sources inside the semi-infinite medium which are in local equilibrium at temperature $T$. 

In global equilibrium the LDOS is given by~\cite{Dorofeyev}
\begin{align}
   D_e^{\rm ge}  (\omega,d)&= \frac{\omega}{\pi c^2} \Im \Tr \bigl[ \mathds{G}^{\rm EE} (\mathbf{r},\mathbf{r}) \bigr] \\
   D_m^{\rm ge} (\omega,d) &= \frac{\mu_0}{\epsilon_0} \frac{\omega}{\pi c^2} \Im \Tr \bigl[ \mathds{G}^{\rm HH} (\mathbf{r},\mathbf{r}) \bigr] 
\end{align}
where  $ \mathds{G}^{\rm HH}$ is the magnetic Green's function. This global equilibrium quantity is typical considered as the LDOS and it coincides with the
local equilibrium expressions for the contribution of the evanescent waves~\cite{Dorofeyev}. Therefore, in the near-field regime where the evanescent waves dominate both definitions of the LDOS are equivalent. Below, we will use the SCUFF-EM package to evaluate $D_e^{\rm ge} (\omega,d)$ and $ D_m^{\rm ge} (\omega,d)$ and compare it to the local equilibrium LDOS $D_e (\omega,d)$ and $D_m (\omega,d)$.

\section{LDOS of sources in a disk}

Now, it can be noted that by definition the local equilibrium LDOS has a very interesting feature. By fixing the volume containing the equilibrated sources at points $\mathbf{r}'$ we can determine how different volume elements are contributing to the LDOS at the position $\mathbf{r}$. In the following, we consider a semi-infinite material with an interface in the x-y plane to vacuum. We want to determine the LDOS at the position $\mathbf{r} = (0,0,d)^t$ generated by the thermal sources contained in a volume of the bulk of the shape of a circular disk with radius $R$ and thickness $t$  (see Fig.~\ref{Fig:sketchvolume}). Then we can do this by simply introducing polar coordinates
\begin{equation}
   \int_V \rd^3 \mathbf{r}' = \int_0^R \rd \rho'\, \rho' \int_0^{2 \pi} \rd \varphi'  \int_{-t}^0 \rd z'.
\end{equation}
But before it is possible to carry out this volume integration, we need the corresponding expressions for the Green functions for point sources within the semi-infinite medium and observation points outside the medium.
These expressions are for a semi-infinite medium well known and can be stated in the Weyl representation as
\begin{equation}
  \mathds{G}^{\rm EE/HH}(\mathbf{r},\mathbf{r}') = \int \!\!\frac{\rd^2 \kappa}{(2 \pi)^2} \,  \re^{\ri \boldsymbol{\kappa}\cdot(\mathbf{x} - \mathbf{x}')} \mathds{G}^{\rm EE/HH} (\boldsymbol{\kappa})
	\label{Eq:Weyl}
\end{equation}
with
\begin{equation}
\begin{split}
  \mathds{G}^{\rm EE} (\boldsymbol{\kappa}) &= \frac{\ri \re^{\ri \gamma_0 z} \re^{-\ri \gamma_1 z'}}{2 \gamma_1} \biggl( t_\rs \mathbf{a}_\rs (\mathbf{k}_1) \otimes \mathbf{a}_\rs (\mathbf{k}_0) \\ &\qquad +  t_\rp \mathbf{a}_\rp (\mathbf{k}_1) \otimes \mathbf{a}_\rp (\mathbf{k}_0) \biggr) 
\end{split}
\end{equation}
and
\begin{equation}
\begin{split}
  \mathds{G}^{\rm HH} (\boldsymbol{\kappa}) &= \frac{1}{c \mu_0} \frac{\ri \re^{\ri \gamma_0 z} \re^{-\ri \gamma_1 z'}}{2 \gamma_1} \biggl( t_\rp \mathbf{a}_\rs (\mathbf{k}_1) \otimes \mathbf{a}_\rs (\mathbf{k}_0) \\ &\qquad +  t_\rs \mathbf{a}_\rp (\mathbf{k}_1) \otimes \mathbf{a}_\rp (\mathbf{k}_0) \biggr).
\end{split}
\end{equation}
Here, we have introduced the wave vector parallel to the interface $\kappa = (k_x,k_y)^t$, the wavevector perpendicular to the interface in vacuum $\gamma_0 = \sqrt{k_0^2 - \kappa^2}$ and within the medium $\gamma_1 = \sqrt{k_0^2 \epsilon_1 - \kappa^2}$ so that $\mathbf{k}_1 = (\kappa, \gamma_1)$ and $\mathbf{k}_0 = (\kappa, \gamma_0)$. Furthermore, we use the notation $\mathbf{x} = (x,y)^t$, $\mathbf{x}' = (x',y')^t$ and the polarization vectors for s- and p-polarization $\mathbf{a}_\rs(\mathbf{k}) = \frac{1}{\kappa} (k_y, - k_x, 0)^t$ and $\mathbf{a}_\rp(\mathbf{k}) = \frac{1}{\kappa k} (- k_x \gamma, - k_y \gamma, \kappa^2)^t$ and the amplitude transmission coefficients
\begin{equation}
  t_\rs = \frac{2 \gamma_1}{\gamma_1 + \gamma_0} \quad \text{and} \quad t_\rp = \frac{2 \sqrt{\epsilon_1} \gamma_1}{\gamma_0 \epsilon_1 + \gamma_1}.
\end{equation}
Note, that $\mathds{G}^{\rm HH}$ can be obtained from $\mathds{G}^{\rm EE}$ by interchanging $t_\rs \leftrightarrow t_\rp$ and multiplying with the factor $1/c \mu_0$. Thus, it suffices to determine $D_e$ because $D_m$ can then be determined from it.

After a lengthy and tedious calculation, we get the final expressions for the local equilibrium LDOS by inserting the expressions for the Green functions into the definition of the LDOS and carrying out the volume integral over a disk-like volume
\begin{widetext}
\begin{equation}
  D_{e/m} = \int_0^\infty \frac{\rd \kappa}{2 \pi}\, \kappa  \int_0^\infty \frac{\rd \kappa'}{2 \pi}\, \kappa' \, \frac{\re^{\ri (\gamma_0 - {\gamma_0'}^*) d}}{4 \gamma_1 {\gamma_1'}^*} \frac{ \gamma_1^2 -  {{\gamma_1'}^*}^2 + \kappa^2 - {\kappa'}^2}{ \gamma_1 - {\gamma_1'}^*} \frac{\omega}{c^2} \biggl(1 - \re^{\ri (\gamma_1 - {\gamma_1'}^*) t} \biggr) \int_0^R \rd \rho' \rho' I_{e/m}(\kappa,\kappa',\rho',\omega)
\label{Eq:LDOSem}
\end{equation}
with the integral kernels
\begin{align}
  I_e(\kappa,\kappa',\rho',\omega) &= \biggl(t_s {t_s'}^* +  t_p {t_p'}^* \frac{\gamma_0 \gamma_1 {\gamma_0'}^* {\gamma_1'}^* }{|k_1|^2 k_0^2} \biggr) C + t_p {t_p'}^* \biggl( \frac{\gamma_1 {\gamma_1'}^* + \gamma_0 {\gamma_0'}^*}{|k_1|^2 k_0^2} \kappa \kappa' B + \frac{\kappa^2 {\kappa'}^2}{|k_1|^2 k_0^2} A  \biggr) \\
  I_m(\kappa,\kappa',\rho',\omega) &=  t_s {t_s'}^* \biggl( \frac{\gamma_0 {\gamma_0'}^*}{k_0^2} C +   \frac{\kappa {\kappa'}^*}{k_0^2} B \biggr) + t_p {t_p'}^*  \biggl( \frac{\gamma_1 {\gamma_1'}^*}{|k_1|^2} A +  \frac{\kappa {\kappa'}^*}{|k_1|^2} B \biggr).
\end{align}
\end{widetext}
We have further introduced the abbreviations 
\begin{align}
  A &= J_0(\rho' \kappa) J_0(\rho' \kappa'), \\
  B &= J_1(\rho' \kappa) J_1(\rho' \kappa'), \\ 
  C &= \frac{1}{2} \bigl( J_0(\rho' \kappa) J_0(\rho' \kappa') + J_2(\rho' \kappa) J_2(\rho' \kappa') \bigr)
\end{align}
where $J_n$ ($n = 0,1,2$) are the cylindrical Bessel functions. Note that the integrals over $\kappa$ and $\kappa'$ are including the propagating waves with $\kappa,\kappa' \leq k_0$ and the evanescent waves for which $\kappa,\kappa' \geq k_0$. 

By taking the limit $t \rightarrow \infty$ and $\rho' \rightarrow \infty$ we have used the relation
\begin{equation}
  \int_0^\infty \rd \rho' \rho' J_{n} (\rho' \kappa)  J_{n} (\rho' \kappa') = \frac{\delta(\kappa - \kappa')}{\kappa}
\end{equation}
to verify that $D_{e/m}$ converges to the well-known expressions for the local equilibrium LDOS of above semi-infinite material~\cite{Dorofeyev}
\begin{equation}
\begin{split}
  D^\infty(\omega,d) &=  D_e^\infty (\omega,d)  +  D_m^\infty (\omega,d) \\
              &= \frac{\omega}{4 \pi^2 c^2} \biggl\{ \int_0^{k_0}\!\!\rd \kappa \, \frac{\kappa}{\gamma_0} \biggl[ (1 - |r_\rs|^2) + (1 - |r_\rp|^2)\biggr]  \\
     &\qquad + \int_{k_0}^\infty\!\!\rd \kappa\, \frac{\kappa^3}{\gamma_0'' k_0^2} \bigl[\Im(r_\rs) + \Im(r_\rp) \bigr] \re^{-2 \gamma_0'' d} \biggr\}
\end{split}
\end{equation}
where $r_{\rs/\rp}$ are the well-known Fresnel reflection coefficiients for s- and p-polarized waves.

\section{Wavevector-cutoff method}

Before comparing the local equilibrium LDOS of the thermal sources inside a disk-shaped volume with the exact LDOS above a finite disk, we want to introduce a simple approximation method for determining the LDOS of a disk-shaped structure. The idea behind this approximation method is rather simple. 
Only waves with a lateral wave vector $\kappa \geq \pi/D$ can contribute to the LDOS of the evanescent field above a disk.
This means the largest allowed wavelength along the interface of the disk is $2 D$ which correspond to a dipolar mode in radial direction of the disk. Then the LDOS due to the evanescent waves can be approximated by 
\begin{equation}
    D(\omega,d) \approx \frac{\omega}{4 \pi^2 c^2} \int_{\pi/D}^\infty\!\!\rd \kappa\, \frac{\kappa^3}{\gamma_0'' k_0^2} \bigl[\Im(r_\rs) + \Im(r_\rp) \bigr] \re^{-2 \gamma_0'' d}. 
\label{Eq:LDOScutoff}
\end{equation}
Since we have here only approximated the evanescent part, this approximation can only be used in the near-field regime where waves with $\kappa > k_0$ give the dominant contribution to the LDOS. Furthermore, this implies that $\pi/D > k_0$ or consequently $D < \lambda/2$ must be fulfilled. Hence, this approximation is only useful in the near-field regime for subwavelength disks.

\begin{figure}
  \includegraphics[width=0.45\textwidth]{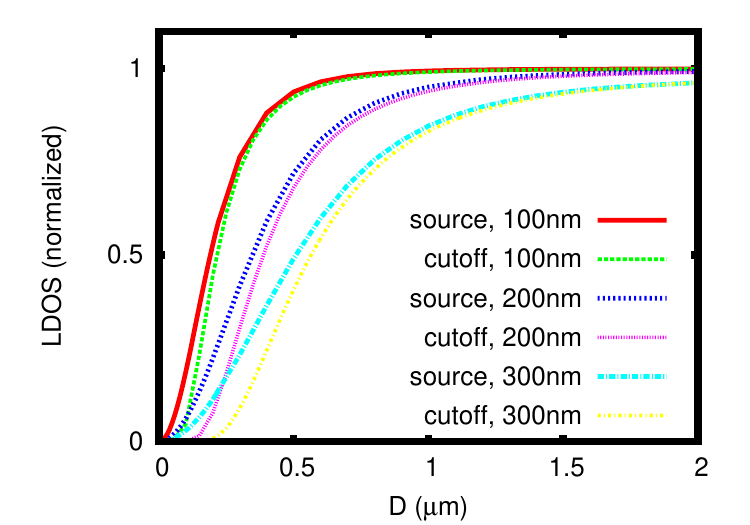}
	\caption{ LDOS at position $\mathbf{r} = (0,0,d)^t$ for a Au disk with diameter $D = 2 R$ at distances $d = 100\,{\rm nm}, 200\,{\rm nm}$, and $300\,{\rm nm}$ using the expressions from Eq.~(\ref{Eq:LDOSem}) labelled as "source" and the cutoff expression from Eq.~(\ref{Eq:LDOScutoff}) labelled as "cutoff". The LDOS is evaluated at a wavelength of $\lambda = 14.5\,\mu{\rm m}$ using the permittivity $\epsilon = -8659 + 4524 \ri$ for Au~\cite{ordal}. The LDOS is further normalized to the LDOS of an Au half space.}
  \label{Fig:VglLDOScutoff}
\end{figure}

In Fig.~\ref{Fig:VglLDOScutoff} we compare the numerical results for the LDOS above an infinitely extended Au half space with sources in an infinitely thick disk, i.e.\ $t \rightarrow \infty$, as function of the disk diameter $D$ with the source method leading to expression~(\ref{Eq:LDOSem}) and the cutoff method in Eq.~(\ref{Eq:LDOScutoff}). It can be seen that the LDOS evaluated with both methods are in good agreement for most values of $D$. For $D \rightarrow 0$ the LDOS goes to zero and converges to the half-space value for $D \rightarrow \infty$. From the source method this behaviour can be understood by the fact that for $D \rightarrow 0$ the source volume of the sources which are generating the LDOS vanishes. For $D \gg d$ enough sources contribute to the LDOS at distance $d$ so that the resulting LDOS coincides with the half-space value. On the other hand, the behaviour can also be understood from the cutoff method. Since the near-field contribution to the LDOS in Eq.~(\ref{Eq:LDOScutoff}) is loosely speaking in the strong near-field regime (where $\gamma_0''\approx \kappa$) dominated by waves with $\kappa \approx 1/d$ we have for $\pi/D \ll 1/d$ or $D \gg \pi d$ no difference to the half-space value. For $D \rightarrow 0$ less and less modes contribute to the $\kappa$ integral in Eq.~(\ref{Eq:LDOScutoff}) and therefore to the LDOS so that it must vanish in this limit. It is interesting that both methods give qualitatively the same result and that for approximately $D > 2 d$ the LDOS determined with both methods also agree quantitatively even though the methods are quite different in their ansatz and their final expressions.

\section{Geometrical vs. thermal boundary conditions}

As is clear from the ansatz of the source method, the LDOS in Eq.~(\ref{Eq:LDOSem}) is the LDOS due to the thermal sources in a disk-shaped part of a semi-infinite sample. It corresponds to the LDOS which can be found above a substrate when it is heated in such a disk-shaped part as depicted in Fig.~\ref{Fig:sketchvolume}. Therefore, the ``boundaries'' set by the finite integration volume can be understood as thermal boundary conditions. When considering the LDOS above a real disk as sketched in Fig.~\ref{Fig:sketch} then the geometrical boundaries due to the finite structure will also play an important role and affect the LDOS above the disk. 

\begin{figure*}
  \includegraphics[width=0.8\textwidth]{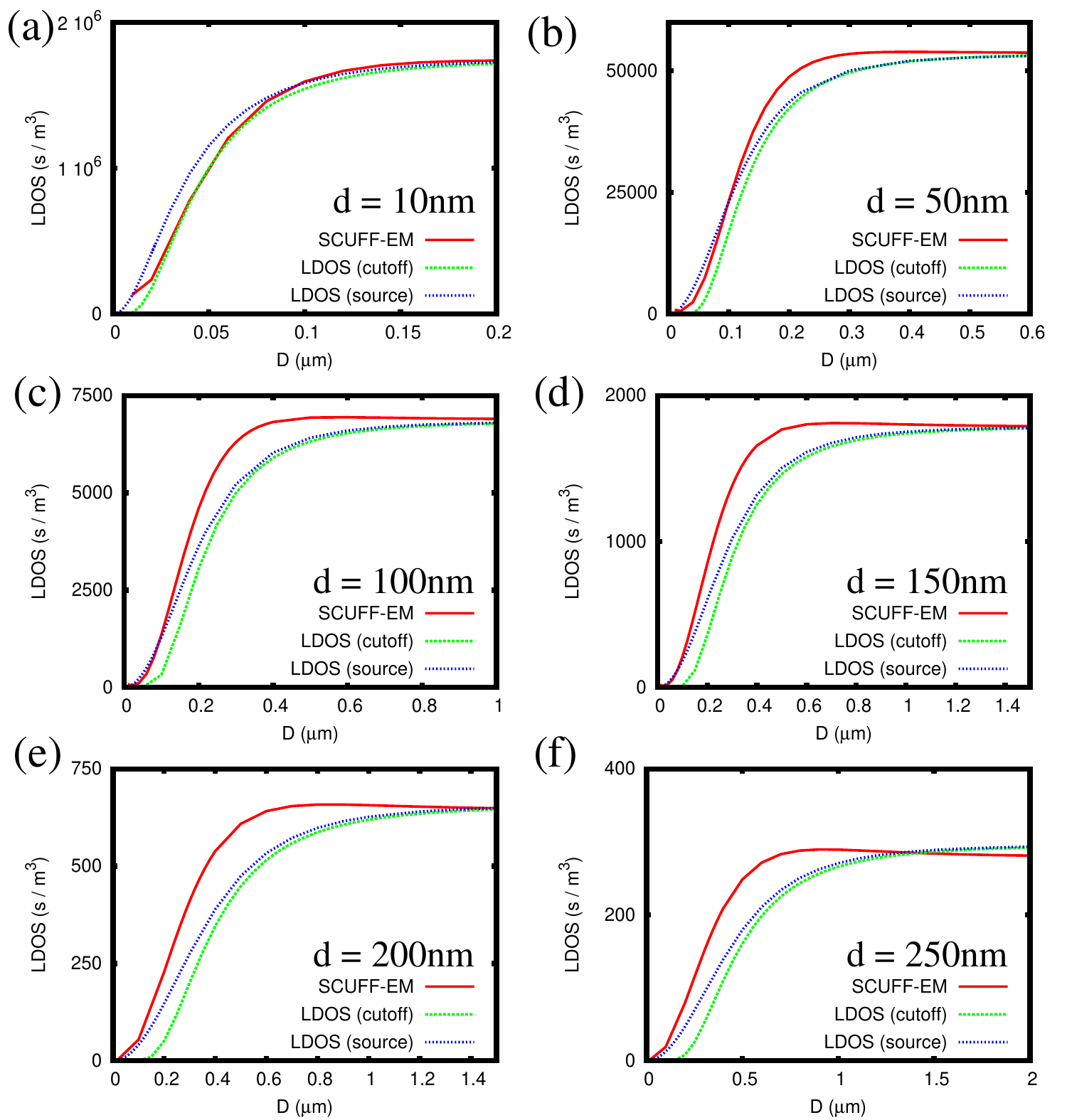}
	\caption{LDOS at position $\mathbf{r} = (0,0,d)^t$ for a Au disk with diameter $D = 2 R$ at distances (a) $d = 10\,{\rm nm}$, (b) $d = 50\,{\rm nm}$, (c) $d = 100\,{\rm nm}$, (d)  $d = 150\,{\rm nm}$,  (e) $d = 200\,{\rm nm}$, (f)  $d = 250\,{\rm nm}$ using the expressions from Eq.~(\ref{Eq:LDOSem}) labelled as "source" and the cutoff expression from Eq.~(\ref{Eq:LDOScutoff}) labelled as "cutoff" as well as the numerically exact SCUFF-EM method. The LDOS is evaluated at a wavelength of $\lambda = 14.5\,\mu{\rm m}$ using the permittivity $\epsilon = -8659 + 4524 \ri$ for Au~\cite{ordal}.}
  \label{Fig:VglLDOSscuffem}
\end{figure*}

\begin{figure}
  \includegraphics[width=0.45\textwidth]{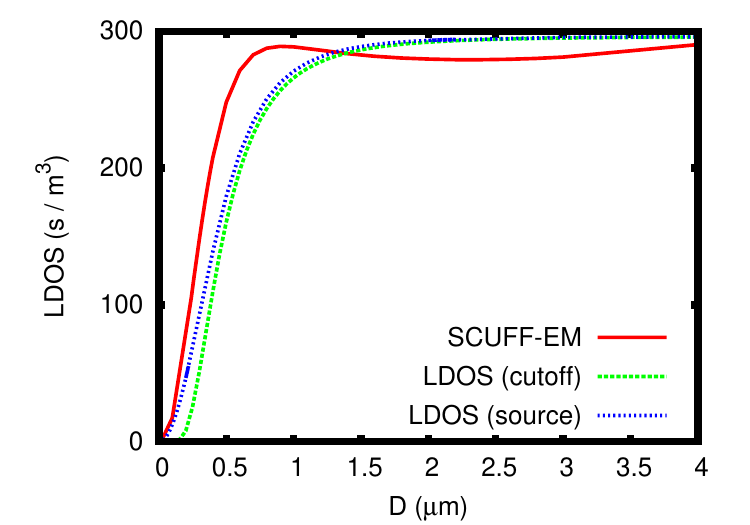}
	\caption{As in Fig.~\ref{Fig:VglLDOSscuffem} for $d = 250\,{\rm nm}$ but for larger range of disk diameter $D$.}
  \label{Fig:VglLDOSscuffem2}
\end{figure}

In Fig.~\ref{Fig:VglLDOSscuffem}, we compare the LDOS obtained with the source and cutoff method with the LDOS as obtained by the exact numerical SCUFF-EM method~\cite{SCUFF1,SCUFF2} for a disk with thickness $t = 300\,{\rm nm}$, so that the disk is much thicker than the skin depth. It can be seen that for very small and very large $D$ the exact LDOS of a metal disk coincides with the LDOS from the source method or the cutoff method, respectively. It is interesting to note that for extremely small distances of only $10\,{\rm nm}$ the LDOS above the disk evaluated with SCUFF-EM coincides more or less with the cutoff method. There is only a small deviation for a diameter of 10~nm which is due to an electric contribution to the LDOS which becomes larger than the magnetic one. The pure magnetic LDOS would for all diameter $D$ coincide with the cutoff method. Consequently, on the one hand for extremely small distances the geometrical boundary conditions tend to make the LDOS smaller than the LDOS due to purely thermal boundary conditions. On the other hand, for intermediate distances and small diameters the source method can better approximate the LDOS of a real finite disk than the cutoff method. In the intermediate regime as for example, for $D \approx 500\,{\rm nm}$ the LDOS above a real nanodisk is larger than above a substrate with a disk-shaped heated region. Hence, one can conclude that here the geometrical boundary conditions are very important in this regime and have the tendency to increase the LDOS. As can be seen for a distance of $250\,{\rm nm}$ this tendency is not very strict, because the exact LDOS seems to overshoot around the LDOS calculated with the source method so that in particular for $d = 250\,{\rm nm}$ the LDOS above a real disk can even be smaller than that above a disk-shaped heated region as seen for $D = 2\,\mu{\rm m}$. Of course, for larger $D$ the exact LDOS will converge to the value of the half space as shown in Fig.~\ref{Fig:VglLDOSscuffem2}. This $D$-dependence in the exact numerical result obtained with SCUFF-EM might be associated to surface plasmon cavity modes of the gold disk.

\section{General temperature profiles}

So far, we have seen that geometrical and thermal boundary conditions for a very simple example can result in similar values for the LDOS. This property can make it difficult in near-field scanning thermal microscopy to disentangle geometrical from thermal information. Here, we want to discuss in more general terms how lateral temperature profiles and surface geometries are related. To keep the discussion simple, we focus on the electrical part of the LDOS or energy density, only. Similar results can of course be obtained for the magnetic LDOS as well.  

On one hand, let us assume that the temperature $T(\mathbf{x})$ of semi-infinite material as depicted in Fig.~1 is a function of the coordinate $\mathbf{x} = (x,y)^t$ which means that we allow a lateral temperature variation in x- and y-direction. Then inserting the Weyl representation of the Green's function in Eq.~(\ref{Eq:Weyl}) into the definition of the electric energy density in Eq.~(\ref{Eq:Ue}) we obtain for the spectral electric energy density
\begin{equation}
	u_E^{T}(\omega) = \frac{\epsilon_0}{2} \int \!\! \frac{\rd \kappa}{(2 \pi)^2} \int \!\! \frac{\rd \kappa'}{(2 \pi)^2} \re^{\ri (\boldsymbol{\kappa} - \boldsymbol{\kappa}')\cdot\mathbf{x}} \tilde{\Theta}(\boldsymbol{\kappa} - \boldsymbol{\kappa}') I_E(\boldsymbol{\kappa},\boldsymbol{\kappa}')
\end{equation}
where
\begin{equation}
	\tilde{\Theta}(\boldsymbol{\kappa} - \boldsymbol{\kappa}') = \int \rd^2 x' \, \Theta(T(\mathbf{x}')) \re^{-\ri (\boldsymbol{\kappa} - \boldsymbol{\kappa}')\cdot\mathbf{x}'}
\end{equation}
is the Fourier transform of the mean energy of a harmonic oscillator $\theta$ for a temperature profile $T(\mathbf{x})$ and $I_E(\boldsymbol{\kappa},\boldsymbol{\kappa}')$ is given by
\begin{equation}
\begin{split}
	I_E (\boldsymbol{\kappa},\boldsymbol{\kappa}') &= \frac{\ri \epsilon_1''}{\gamma_1 - {\gamma_1'}^*}\frac{\re^{\ri (\gamma_0 - {\gamma_0'}^*) z}}{4 \gamma_1 {\gamma_1'}^*} \biggl[ t_\rs^{12} {{t_\rs^{12}}'}^* \frac{(\boldsymbol{\kappa} \cdot\boldsymbol{\kappa}')^2}{\kappa^2 {\kappa'}^2} \\
	 &\quad +  t_\rp^{12} {{t_\rp^{12}}'}^* \frac{\gamma_0 {\gamma_0'}^* (\boldsymbol{\kappa} \cdot\boldsymbol{\kappa}') + \kappa^2 {\kappa'}^2}{\kappa {\kappa'} k_0^2} \\
	 &\quad\times\frac{\gamma_1 {\gamma_1'}^* (\boldsymbol{\kappa} \cdot\boldsymbol{\kappa}') + \kappa^2 {\kappa'}^2}{\kappa {\kappa'} |k_1|^2} \biggr].
\end{split}
\end{equation}
This result allows for determining the LDOS for any lateral temperature variation assuming that variations in z-direction within the material can be neglected.

On the other hand, now we assume that the surface of the semi-infinite medium is not flat, but is described by a surface profile function $h f(\mathbf{x})$ where $h$ is the height parameter of the profile. Then a perturbation approach in first order of the surface profile height $h$ yields the first-order correction to the LDOS of a flat surface of the form~\cite{LDOSProfile}
\begin{equation}
\begin{split}
	u_E^{f}(\omega) &=  \frac{\epsilon_0}{2} \Theta(T) \Im \biggl( \frac{1 - \epsilon_1}{\omega \pi}\int \!\! \frac{\rd \kappa}{(2 \pi)^2} \int \!\! \frac{\rd \kappa'}{(2 \pi)^2} \\
	&\quad \times \re^{\ri (\boldsymbol{\kappa} - \boldsymbol{\kappa}')\cdot\mathbf{x}} \tilde{f}(\boldsymbol{\kappa} - \boldsymbol{\kappa}') \tilde{I}_E(\boldsymbol{\kappa},\boldsymbol{\kappa}') \biggr)
\end{split}
\end{equation}
where
\begin{equation}
	\tilde{f}(\boldsymbol{\kappa} - \boldsymbol{\kappa}') = h \int \rd^2 x' \, f(\mathbf{x}') \re^{-\ri (\boldsymbol{\kappa} - \boldsymbol{\kappa}')\cdot\mathbf{x}'}
\end{equation}
is the Fourier transform of the surface profile. The function $\tilde{I}_E$ is defined in Ref.~\cite{LDOSProfile} and resembles $I_E$. 

Now, we can compare this expression $u_E^f$ with $u_E^T$ and first of all note that both expressions cannot be directly related to each other. However, in both cases the resulting lateral variation of the energy density due to a temperature or surface profile results in a convolution of the similar functions $I_E$ and $\tilde{I}_E$ with the Fourier transform of the surface profile function $h f(\mathbf{x})$ or the temperature profile function $\Theta(T(\mathbf{x}))$. 
This formal resemblance underlines the finding that thermal and geometrical boundary conditions have a similar impact. This might make it difficult to distinguish if a measured signal is due to heating or geometrical effects so that further theoretical and experimental investigations are needed to disentangle between thermal and geometrical boundary conditions. Also temperature variations in $z$-direction as considered for near-field thermal radiation~\cite{RM1,RM2,RM3} need to be taken into account in future works.

\section{Conclusions}

We have determined the analytical expression for the LDOS above a disk-shaped patch within a semi-infinite sample. The such determined LDOS is due to the thermal sources in this disk-like patch of the semi-infinite structure. We have further introduced a simple cutoff-method which gives very similar results as the source method. We found by numerical comparison that for a gold disk the results for both methods agree very well for disks with a diameter of $D > 2 d$. Finally, we have compared the results of the source method with the exact LDOS above a metal disk using SCUFF-EM. This allows us to discuss the difference between thermal and geometrical boundary conditions. We have shown that due to the geometrical boundary conditions the exact LDOS of an Au disk typically overshoots the values of the LDOS obtained with the source method. For extremely small distances of 10~nm, the geometrical boundary condition have the tendency to give an LDOS which is smaller than obtained with the source method and which is very well described by the cutoff method. For intermediate distances between $100\,{\rm nm} - 200\,{\rm nm}$ the geometrical boundary conditions have the tendency to enhance the LDOS compared to the LDOS obtained from the source method, even though for very small and large $D$ the exact LDOS coincides with the LDOS obtained with the source method. We expect that similar results can be found for polar materials. However, since such materials have resonances in the infrared, a detailed study for a broad range of frequencies around these resonances is necessary which is out of the scope of our work. 

Hence, we conclude that in a LDOS measurement with instruments like the NSthM and the SNoiM thermal and geometrical inhomogeneities will lead to similar signals so that a clear distinction between geometrically and thermally induced effects might not always be possible. This observation is particularly important in experiments studying samples that possess nanostructures and temperature distributions. As pointed out by our results from theory for such experimental setups it is of utmost importance for a correct interpretation to use combined measurement methods or information channels to disentangle the geometrical and thermal information. Obviously, with near-field scanning thermal microscopes based on AFMs or STMs, like the NSThM and TINS, it is possible to obtain a thermo-signal and a signal which contains the geometric information due to the AFM and STM capabilities. In this case, a comparison of theoretical results using the geometric information and the experimental thermosignal will allow to disentangle both informations in the thermosignal. For microscopes like the TRSTM and SNoiM one could use samples which are precharacterized with an AFM or STM. Another method could be to make measurements of the thermosignal at two different wavelengths $\lambda_1$ and $\lambda_2$ with $|\lambda_1 - \lambda_2|$ much smaller than the Planck window. The intensity contrast in this case only contains the signal due to variation of the geometry or material properties in the case of inhomogeneous samples. Finally, TRSTM and SNoiM measurements with two microscope probes of different materials could be a method to disentangle the geometric and thermal signal. Clearly, further theoretical and experimental works are necessary to quantify and pinpoint the impact of thermal and geometrical inhomogeneities.

\acknowledgments
\noindent

S.-A.\ B. acknowledges support from Heisenberg Programme of the Deutsche Forschungsgemeinschaft (DFG, German Research Foundation) under the project No. 461632548 and support from the QuantUM program of the University of Montpellier, and he thanks the University of Montpellier and the group Theory of Light-Matter and Quantum Phenomena of the Laboratoire Charles Coulomb for hospitality during his stay in Montpellier where part of this work has been done.   Z.A. acknowledges support from National Natural Science Foundation of China (NSFC) under the project No.\ 12027805 and Shanghai Science and Technology Committee under project No.\ 20JC1414700. The authors further acknowledge support from the Sino-German Center for Research Promotion (No.\ M-0174). This work has received funding from the European Community through the Horizon 2020 research and innovation programs under grant agreement No. 766853 (EFINED).

\end{document}